\documentclass[12pt]{iopart}
\usepackage{iopams} 
\usepackage{graphicx}
\usepackage[T1]{fontenc}

\newcommand{\al}{\alpha}
\newcommand{\bta}{\beta}
\newcommand{\gm}{\gamma}
\newcommand{\lb}{\lambda}
\newcommand{\eps}{\epsilon}
\newcommand{\vth}{\vartheta}

\newcommand{\om}{\omega}
\newcommand{\Om}{\Omega}
\newcommand{\veps}{\varepsilon}

\newcommand{\beq}{\begin{equation}}
\newcommand{\eeq}{\end{equation}}
\newcommand{\ba}{\begin{array}}
\newcommand{\ea}{\end{array}}
\newcommand{\bn}{\begin{eqnarray}}
\newcommand{\en}{\end{eqnarray}}
\newcommand{\bc}{\begin{center}}
\newcommand{\ec}{\end{center}}
\newcommand{\bit}{\begin{itemize}}
\newcommand{\eit}{\end{itemize}}
\newcommand{\btab}{\begin{tabular}}
\newcommand{\etab}{\end{tabular}}
\newcommand{\half}{\frac{1}{2}}

\newcommand{\thalf}{\case{1}{2}}

\newcommand{\bls}[1]{\boldsymbol{#1}}
\newcommand{\mt}[1]{\mathsf{#1}}

\eqnobysec

\begin{document}
\title[GOA for the quadrupole states]{Gaussian Overlap Approximation for the quadrupole collective states}

\author{ Stanis{\l}aw G Rohozi\'nski}

\address{ Institute of Theoretical Physics, Faculty of Physics, University of Warsaw, \\ Ho\.za 69, 00-681 Warszawa,  Poland}
\ead{Stanislaw-G.Rohozinski@fuw.edu.pl}

\begin{abstract}
The Generator Coordinate Method (GCM) in the Gaussian Overlap Approximation (GOA) is applied to a description of the nuclear quadrupole collective
states. The full five-dimensional quadrupole tensor is used as a set of the generator coordinates. The integral Hill-Wheeler equation is reduced
to a differential equation by using the Fourier transforms of the overlap and energy kernels. The differential Bohr Hamiltonian obtained this way is compared with that derived by the usual approach to the collective Hamiltonian in the GOA which does contain an additional approximation. The method of calculating the quantities which determine the Bohr Hamiltonian from the set of deformation-dependent intrinsic states is demonstrated.
In particular, it appears that the moments of inertia at the quadrupole rotations are of the type of that of Yoccoz.   
\end{abstract}

\pacs{21.60.Ev, 21.60.Jz, 21.10.Re, 21.10.Ky}
\submitto{\jpg}
\maketitle

\section{Introduction}\label{I}

The Bohr Hamiltonian (cf a recent review \cite{Pro09}) still remains one of the most effective tools to describe the lowest 
collective quadrupole excitations of the even-even nuclei.
This is in spite of a long history of the collective model dated from the fifties of the twentieth century \cite{Boh52}. The rudiments of the model
can be met even earlier \cite{Flu41}. 
Nowadays the Bohr Hamiltonian in a general form proposed in \cite{Bel65,Kum67} is considered. It is based on the correspondence principle,
that is, on the assumption that the collective quadrupole motion has its classical counterpart. The quantum Bohr Hamiltonian is constructed from the 
most general classical Hamiltonian quadratic in velocities by the Podolsky-Pauli quantization prescription \cite{Pod28,Pau33,Hof72}. The quantum 
kinetic energy operator is then the Laplace-Beltrami operator in the Riemannian space with the metric tensor equal to the classical inertial matrix.  Derivations of the Bohr Hamiltonian from the microscopic many-body theories of nuclei used so far, which are based on methods of 
the type of the Time-Dependent Hartree-Fock method, do pass through their classical stage too. The collective motion is then represented by a wave packet, which keeps the classical law of motion, requantized afterwards (cf e.g. \cite{Bar68,Pro09}).

A possibility of a purely quantum description of  collective excitations is created by the Generator Coordinate Method (GCM) \cite{Hill53,Gri57}
(see also \cite{Rin80}).
The method leads originally to an integral eigenvalue equation (the Hill-Wheeler equation).  
The integral equation can be approximated by a differential equation of the Sch\"odinger type \cite{Haf73}.
It is especially easy to do when the Gaussian Overlap Approximation (GOA) is made for overlap and energy kernels in the Hill-Wheeler equation
\cite{Gri57,Jan64,Bri68}. 
The method is usually demonstrated for a single generator coordinate. But the quadrupole motion has five degrees of freedom and thus five generator coordinates are required. The multi-dimensional GCM was already considered by several authors \cite{Kam73,Oni75,Goz85a,Goz85b} (see also the review article \cite{Rei87}).
Obviously, the GCM, being one of the contemporary tools for investigations of various collective phenomena and for the restoration of broken symmetries, 
has been applied to various problems in the nuclear physics apart from the quadrupole collective motion. The two-dimensional Hill-Wheeler equation
has been used to investigate the quadrupole $\bta$ and $\gm$ vibrations \cite{Bon90,Bon91,Hee93}. Before that  the differential equations coming from the GOA have been applied to these  vibrations \cite{Rei78,Ner86,Sta89,Pil92}. The GOA has been used to estimate the ground-state correlations associated with the quadrupole collective modes \cite{Hag02}. But the rotational correction to the ground-state energy has been obtained through the projection onto the 
good-angular-momentum state. The five-dimensional collective quadrupole Hamiltonian  
based on the GOA has been generated from the microscopic theory with the Gogny effective forces and used in comprehensive calculations of the $2^+$ nuclear excitations \cite{Ber07}. However, the moments of inertia obtained by the Thouless-Valatin prescription \cite{Tho62} instead of those derived from the GOA are used in the Bohr Hamiltonian. In spite of these developments, it seems there is no consistent derivation of the full five-dimensional Bohr Hamiltonian using the GCM with the GOA yet.

The present study is devoted to application of the multi-dimensional GCM in the GOA to the five-dimensional quadrupole collective motion of the
even-even nuclei. The quadrupole deformation tensor plays a role of a set of the five generator coordinates. We aim at deriving the Bohr Hamiltonian
in a purely quantum way. The Hamiltonian obtained in such a way is a bit more general than that based on the correspondence principle (cf \cite{Pro09}).
In order to transform the Hill-Wheeler integral eigenvalue equation into the Schr{\"o}dinger differential equation with the Bohr Hamiltonian we introduce, 
after Onishi and Une \cite{Oni75}, new quadrupole variables which diagonalise the exponent of the Gaussian overlap. Using these new variables we perform  
the Fourier analysis of the energy kernel. The kernel turns out to be a local distribution containing the Dirac delta function and its derivatives of the second order. It allows us to reduce the integral equation to a differential one.  The Hamiltonian of the rigid rotor was derived earlier using the same method \cite{Une76}. 
We compare the method of the Fourier analysis with the method used usually in deriving the collective Hamiltonian from the GCM  
\cite{Rin80,Rei87,Goz85b}. The latter method contains an additional approximation apart from the GOA.

We formulate the Hill-Wheeler theory for the quadrupole collective excitations in \sref{HW}. \Sref{GOA} is devoted to the Gaussian Overlap Approximation for the quadrupole generator coordinates. The differential equation equivalent to the Hill-Wheeler equation in the GOA is derived and compared with the usual collective approach to the GCM in \sref{DHW}. Formulae for calculating the quantities determining the overlaps within the GOA and the Bohr Hamiltonian from the deformation-dependent intrinsic state are given in \sref{CK}. We draw conclusions from the present study in \sref{C}. In \ref{SCWF} the semi-Cartesian Wigner functions are introduced, which are convenient to use in the present study. The structure of matrices being isotropic functions of the quadrupole variables is discussed in \ref{QM}.

\section{The Hill-Wheeler theory of the quadrupole collective excitations}\label{HW}

\subsection{The quadrupole generating states}\label{GS}
Let the considered nucleus with $Z$ protons and $N$ neutrons be described microscopically by the many -body Hamiltonian $\hat{H}$ which is invariant under the $O^T$(3) group of transformations i.e. the superposition of time reversal $\hat{T}$ and the orthogonal transformations (the rotations $\hat{R}(\omega )$ and the space inversion $\hat{P}$) in the physical three-dimensional space.
In order to describe its quadrupole collective excitations by means of the GCM we start with states parametrised by  
the two deformation parameters $d_0$ and $d_2$ which describe the quadrupole deformation with respect to a system of principal axes. For instance, the intrinsic components of the mass quadrupole moment in the deformed ground state can serve as such parameters. Parameters $d_0$ and $d_2$ define the deformation 
(e.g., a shape) up to the group of five permutations of the principal axes (cf \cite{Pro09}). One of the principal axes is the symmetry axis when one of the three following conditions for the deformation parameters is fulfilled:
$d_2=0$, $d_2=\pm\sqrt{3}d_0$. The conditions represent the three straight lines (or the six rays) intersecting each other at point $d_0=0,\ d_2=0$ on plane
$(d_0 d_2)$ of the deformation parameters. The rays divide the plane into the six sectors. The values of $d_0$ and $d_2$ located in one of the sectors describe
the all possible deformations (shapes) at the established names of principal axes. The standard choice of the sector is the following:
\beq\label{drange}
0\leq d_2\leq \sqrt{3}d_0<+\infty  ,
\eeq
which means that  the symmetry axes are  $z$ and  $y$ for $d_2=0$  and $d_2=\sqrt{3}d_0$, respectively.
Here, we assume that the ranges of values of $d_0$ and $d_2$ are given also by \eref{drange}. 
We denote for short  the set  $(d_0,\ d_2)$ as $d$. 
Let $|\phi (d)\rangle$ be an approximated ground state  of the considered nucleus with a given deformation 
$d$. State $|\phi (d)\rangle$ is supposed to be the ground state of the mean-field Hamiltonian, either the phenomenological or the self-consistent one with constraints for the quadrupole moment. 
For all possible values of $d$ states $|\phi (d)\rangle$ form a set of states called the intrinsic states in the following. 
The normalization of the intrinsic state is assumed to be:
\beq\label{norm}
\langle\phi (d)|\phi (d)\rangle =1
\eeq
for every $d$. 

In general, the state $|\phi (d)\rangle$ is not an irreducible representation of $O^T$(3) because the rotational symmetry is usually broken spontaneously in the mean field. 
However, it is assumed that 
$|\phi (d)\rangle$ is still invariant under the point $D^T_{2h}$ group \cite{Dob00} with the intrinsic axes $(x,\ y,\ z)$ as its symmetry axes and we have:
\numparts
\bn
&&\hat{T}|\phi (d)\rangle =|\phi (d)\rangle ,\label{timerev} \\
&&e^{-\rmi\pi \hat{J}_u}|\phi (d)\rangle =|\phi (d)\rangle ,\label{signature} \\
&&\hat{P}|\phi (d)\rangle =|\phi (d)\rangle , \label{parity} 
\en
\endnumparts
where $\hat{J}_u$ for $u=x,y,z$ are the intrinsic components of the total angular momentum operator. 
\Eref{timerev} can be fulfilled only for even $Z$ and $N$ and, as a matter of fact, it corresponds to the assumption 
that we will consider only the even-even nuclei here. Although the $D^T_{2h}$
symmetry looks natural, the parity invariance \eref{parity} will not be used explicitly in further considerations. 
It is because the quadrupole deformation tensor itself is of  the positive parity.
Since the differentiation with respect to the deformation parameters $d$ commutes with the $D^T_{2h}$ generators the states 
\beq\label{deriv}
|\phi_k(d)\rangle = \frac{\partial}{\partial d_k}|\phi (d)\rangle ,\quad 
|\phi_{kl}(d)\rangle = \frac{\partial^2}{\partial d_k\partial d_l}|\phi (d)\rangle ,\quad\dots
\eeq
for $k,\ l=0,\ 2$ have also the $D^T_{2h}$ symmetry and fulfil \eref{timerev}, \eref{signature} and \eref{parity}.

It follows from time reversal invariance  that the matrix elements obey relations
\bn\label{timeev}
&&\langle \phi (d)|\phi (d')\rangle = \langle \phi (d)|\phi (d')\rangle^\ast =\langle \phi (d')|\phi (d)\rangle ,\nonumber \\
&&\langle \phi (d)|\hat{H}|\phi (d')\rangle = \langle \phi (d)|\hat{H}|\phi (d')\rangle^\ast =\langle \phi (d')|\hat{H}|\phi (d)\rangle ,
\en
i.e. they are real and symmetric with respect to $d$ and $d'$. On the other hand, matrix elements 
$\langle \phi (d)|\hat{J}_u|\phi (d')\rangle$ and $\langle \phi (d)|\hat{J}_u\hat{H}|\phi (d')\rangle$ are imaginary and antisymmetric, because 
the angular momentum operators $\hat{J}_u$ are time odd.

It follows from the $D_{2}$ symmetry \eref{signature} of $|\phi (d)\rangle$ and  $|\phi_i (d)\rangle$ that, in particular,
 the matrix elements of type
$\langle\phi_i(d)|\hat{J}_u|\phi (d)\rangle$, $\langle\phi_i(d)|\hat{J}_u\hat{H}|\phi (d)\rangle$,  
and $\langle\phi (d)|\hat{J}_u\hat{J}_{u'}|\phi (d)\rangle$, $\langle\phi (d)|\hat{J}_u\hat{J}_{u'}\hat{H}|\phi (d)\rangle$ for $u\ne u'$ all vanish. 

Since $|\phi (d)\rangle$ does not possess the rotational symmetry, the state rotated by the three Euler angles 
$\omega =(\om_1,\om_2,\om_3)$ from the ranges 
$0\leq\om_1<2\pi$, $0\leq\om_2\leq\pi$, $0\leq\om_3<2\pi$
\beq\label{rotstate}
 |\Phi (d,\omega )\rangle = \hat{R}(\omega )|\phi (d)\rangle 
\eeq
is an equally good ground state of the mean-field Hamiltonian. The unitarity of $\hat{R}(\omega )$ assures us that it is also normalised  to 1. 
Operator $\hat{R}(\omega )$ is time-even
and thus the  matrix elements between states $|\Phi (d,\om )\rangle$ and $|\Phi (d',\om ')\rangle$ corresponding
to these of \eref{timeev} are also real and symmetric with respect to sets of variables $(d,\om )$ and $(d',\om ')$.
States \eref{rotstate}
will play the role of the generating states  and the five variables $d_0$, $d_2$, 
$\om_1$, $\om_2$ and $\om_3$ can be used as the generator coordinates for the quadrupole collective excitations.
As a matter of fact, other generator coordinates, namely the laboratory quadrupole deformations defined below will be used.

\subsection{Quadrupole coordinates}\label{QCO}
We introduce a new set of the five variables, namely $\al_\mu$, $\mu =-2,\dots ,+2$, by the relation
\beq\label{alb}
\al_\mu = d_0D_{\mu 0}(\omega ) + d_2D_{\mu 2}(\omega ) ,
\eeq
where the semi-Cartesian Wigner functions $D_{\mu 0}$ and $D_{\mu 2}$, defined by \eref{WCart}, are combinations of the usual Wigner functions $\mathcal{D}^2_{\mu\nu}$. They play the role of the laboratory deformation parameters. The Jacobian of transformation 
\eref{alb} is equal to (cf \cite{Pro09})
\beq\label{jacob}
W(d,\om )=|d_2(3d_0^2-d_2^2)|\sin{\om_2}
\eeq
and, therefore, the transformation is reversible in the entire assumed open ranges of variables $d$ and $\om$.
It is seen from \eref{alb} that $\al_\mu$ are the components of an electric (i.e. with positive parity), real
(i.e. $\al_\mu^\ast = (-1)^\mu\al_{-\mu}$) quadrupole tensor $\bls{\al}$ in the three-dimensional physical space. 
The complex conjugate components are denoted with the superscript, e.g.
\beq\label{complex}
\al^\mu =\al_\mu^\ast =(-1)^\mu\al_{-\mu}.
\eeq
The frame of axes with orientation $\om =0$ is the frame of principal axes (intrinsic frame) of $\bls{\al}$, because the intrinsic components of $\bls{\al}$ are equal to $\al_0(\om =0) =d_0$, $\al_2(\om =0) =\al_{-2}(\om =0)=d_2/\sqrt{2}$ and $\al_{1}(\om =0)=\al_{-1}(\om =0)=0$.

Instead of the complex spherical components of the tensor, the five truly real coordinates 
$a_k$, $k =0, \ 2,\ x,\ y,\ z$, namely 
\beq\label{reco}
a_k=D^\mu_k(\om =0)\al_\mu =C^\mu_k\al_\mu
\eeq
can be used, where the Einstein contraction rule is applied to summing of the Greek upper and lower indeces $\mu =-2,\ \dots ,+2$ of the spherical components. The rule is not applied to the Latin Cartesian indeces. These coordinates can be treated as the Cartesian components of vector $\bi{a}$ in the five-dimensional space. 
This is because the scalar product of the two quadrupole tensors, $\bls{\al}$ and $\bls{\bta}$ is
\beq\label{scpr}
\bls{\al}\cdot\bls{\bta}=\sum_{\mu}(-1)^\mu\al_\mu\bta_{-\mu} =\al_\mu\bta^\mu =\sum_{k}a_kb_k =\bi{a}\cdot\bi{b} .
\eeq
In connexion with \eref{scpr} the volume element in the space of the quadrupole coordinates is:
\beq\label{vol}
\mathrm{d}\Omega (\bls{\al})=\Pi_k\mathrm{d} a_k =|d_2(3d_0^2-d_2^2)|\sin{\om_2}\rmd d_0\rmd d_2\rmd\om_1\rmd\om_2\rmd\om_3 .
\eeq 

All functions of deformation $d$ and the Euler angles, $\om$, can be treated as functions of either tensor $\bls{\al}$ or vector $\bi{a}$.
Here, we prefer to use the complex variables, $\al_\mu$, for convenience. However, we have in mind that the integration variables are always 
real: either $a_k$ or $d$ and $\om$ as is seen in \eref{vol}.
The relation between the derivatives with respect to $\al_\mu$ and the derivatives with respect to the Euler angles and the intrinsic components 
$d_0$, $d_2$ reads \cite{Roh82,Pro09}:
\bn\label{dalpha}
\frac{\partial}{\partial\al_\mu}&=& \sum_kC^\mu_k\frac{\partial}{\partial a_k} \nonumber \\
&=&D^\mu_0(\omega )\frac{\partial}{\partial d_0}
+D^{\mu}_{2}(\omega )\frac{\partial}{\partial d_2} 
 +\rmi D^{\mu}_{z}(\omega )\frac{1}{2d_z}L_z(\omega ) \nonumber \\
&&-\rmi D^{\mu}_{x}(\omega )\frac{1}{2d_x}L_x(\omega )
-\rmi D^{\mu}_{y}(\omega )\frac{1}{2d_y}L_y(\omega ),
\en
where
\beq\label{dxyz}
d_x=-\frac{1}{2}(\sqrt{3}d_0+d_2),\quad d_y=\frac{1}{2}(\sqrt{3}d_0-d_2),\quad d_z=d_2 
\eeq       
and $L_x(\omega ),\ L_y(\omega ),\ L_z(\omega )$ are differential operators with respect to the Euler angles and are interpreted as 
the intrinsic components of the drift angular momentum of the rotation of the intrinsic frame (eq. (2.15) in ref. \cite{Pro09}).

\subsection{Trial states for the collective quadrupole excitations}\label{TS}
We denote the generating states \eref{rotstate} as $|\Phi (\bls{\al})\rangle$. Then
trial states for the quadrupole collective motion are taken in the form:
\beq\label{varst}
|\Psi [\varphi ]\rangle =\int\varphi (\bls{\al})|\Phi (\bls{\alpha})\rangle\rmd\Omega (\bls{\al}).
\eeq
The five components of the quadrupole tensor $\bls{\al}$ play the role of the generator coordinates and function $\varphi (\bls{\al})$ is 
the weight function. The variational principle leads to the Hill-Wheeler
integral equation of the form
\beq\label{hw}
\int[\mathcal{H}(\bls{\al},\bls{\al}^{\prime})- E\mathcal{I}(\bls{\al},\bls{\al}^{\prime})]\varphi (\bls{\al}^{\prime})\rmd\Omega (\bls{\al}^\prime ) =0
\eeq
for function $\varphi$ of the generator coordinates.
The equation is determined by the two real symmetric kernels: the overlap kernel 
\beq\label{overlap}
\mathcal{I}(\bls{\al},\bls{\al}^{\prime})=\langle\Phi (\bls{\al})|\Phi (\bls{\al}^{\prime})\rangle
\eeq
and the energy kernel
\beq\label{energy}
\mathcal{H}(\bls{\al},\bls{\al}^{\prime})=\langle\Phi (\bls{\al})|\hat{H}|\Phi (\bls{\al}^{\prime})\rangle.
\eeq
The overlap kernel \eref{overlap} is non-diagonal and equation \eref{hw} for the weight functions constitutes 
a non-orthogonal eigenvalue equation.

\section{The Gaussian Overlap Approximation}\label{GOA}
\subsection{Gaussian approximation for the kernels}\label{GAK}
We expect that the overlap kernel drops quickly from one to zero when  differences $\gm_\mu =\al_\mu -\al^{\prime}_\mu$ of the components of tensors 
$\bls{\al}$ and $\bls{\al}^{\prime}$ increase from zero. Therefore, the logarithm of the kernel rather than the kernel itself  as a function 
of $\bls{\gm}$ can be approximated by the power series:
\beq\label{lnover}
\ln{\mathcal{I}(\bls{\bta}+\thalf\bls{\gm},\bls{\bta}-\thalf\bls{\gm})}= -\half g^{\mu\nu}(\bls{\bta})\gm_\mu\gm_\nu +\dots ,
\eeq
where $\bls{\bta}=\thalf (\bls{\al}+\bls{\al}^{\prime})$ and the Einstein contraction rule is applied to upper and lower indeces $\mu,\ \nu$. When in expansion \eref{lnover} we keep
only the term quadratic in $\bls{\gm}$ for an arbitrary  $\bls{\bta}$,  we have the local Gaussian approximation for the overlap kernel:
\beq\label{gaussover}
\mathcal{I}(\bls{\bta}+\thalf\bls{\gm},\bls{\bta}-\thalf\bls{\gm})=\exp{(-\half g^{\mu\nu}(\bls{\bta})\gm_\mu\gm_\nu)}.
\eeq
Matrix $\mt{g}$ is real, symmetric and  positive definite.
We use a similar approximation for quotient
\beq\label{gaussen}
 \frac{\mathcal{H}(\bls{\bta}+\thalf\bls{\gm},\bls{\bta}-\thalf\bls{\gm})}
{\mathcal{I}(\bls{\bta}+\thalf\bls{\gm},\bls{\bta}-\thalf\bls{\gm})}=v(\bls{\bta})-\half h^{\mu\nu}(\bls{\bta})\gm_{\mu}\gm_{\nu}
+\dots .
\eeq
Matrix $\mt{h}$ is real and symmetric. 

Both matrices, $g_{\mu\nu}$ and $h_{\mu\nu}$, are
the symmetric quadrupole bitensors in the interpretation of \ref{QM}, whereas $v$ a scalar.  
We shall refer to matrix $\mt{g}$ as the metric tensor. Throughout the paper the upper and lower indices of matrices are connected with
the complex conjugation of the spherical tensors according to \eref{complex}, and not with the contravariant and covariant components of tensors in the Riemannian space. A separate symbol is used for the inverse metric tensor, namely $\mt{g}^{-1}=\mt{f}$.
  
\subsection{Eigenvalue equation for the Gaussian overlap kernel}\label{EVGO}

To find the eigenfunctions of the Gaussian overlap kernel \eref{gaussover} it is convenient to reduce the exponent of the Gaussian function to 
the sum of squares of five variables. It is a simple thing
to do for the single generator coordinate \cite{Fer71}. It is not so simple in the case of a set of a few variables. Here, we adopt the defininion of the new variables given by Onishi and Une \cite{Oni75}. Namely,
we introduce  the five new real variables  $t_k$ ($k=$0, 2, x, y, z) through 
a line integral in the five-dimensional space
of vectors $\bi{a}$ of \eref{reco}
\beq\label{nvar}
t_k(\bi{a})=\int^{\bi{a}}\sum_l r_{kl}(\bi{a}^{\prime})\rmd a_{l}^{\prime}.
\eeq
The lower integration limit is omitted in the notation of the integral because it is not relevant here.
Matrix $\mt{r}$ is taken to be the positive-definite symmetric square root matrix of $\mt{g}$ defined by 
\eref{sqmatintr} with $\mt{m}=\mt{g}$ and $\mt{m}^{1/2}= \mt{r}$. It fulfils relation \eref{sqmat}, namely
\beq\label{sqm}
g_{kl}(\bi{a})=\sum_ir_{ki}(\bi{a})r_{il}(\bi{a}),
\eeq
where $g_{kl}$ are the Cartesian components of $\mt{g}$ defined according to \eref{cart}.
The variables $t_k$  are defined well if the line integral does not depend on the path of integration. It is so for the irrotational field 
$r_{kl}(\bi{a})$ i.e. for
\beq\label{wl}
\frac{\partial r_{kl}}{\partial a_i}-\frac{\partial r_{ki}}{\partial a_l}=0 ,
\eeq
which we assume for use of further considerations. However, we do not know in advance whether it is really fulfilled.

It is convenient to operate with the complex linear combinations $\tau_\mu$ of $t_k$ defined according to \eref{reco}. Then we have
\beq\label{tcom} 
\rmd\tau_\mu (\bls{\al})=\sum_{k,l}C_{\mu k}C^\nu_l r_{kl}(\bi{a})\rmd a_l =r_\mu^\nu (\bls{\al})\rmd\al_\nu ,
\eeq
where $r_\mu^\nu =r_\mu^{\phantom{\mu}\nu}=r_{\phantom{\nu}\mu}^\nu$ are the corresponding spherical components of $\mt{r}$ (the sequence of the superscript and subscript does not need to be fixed for a symmetric matrix).
Relations \eref{sqm} and \eref{wl} have the following counterparts for the spherical components:
\bn\label{sqmwl}
\fl g_{\mu\nu}(\bls{\al})-r_{\mu}^{\lb}(\bls{\al})r_{\lb\nu}(\bls{\al})
=\sum_{k,l}\left(C_{\mu k}C_{\nu l}\left(g_{kl}(\bi{a})-\sum_ir_{ki}(\bi{a})r_{il}(\bi{a})\right)\right) =0, \\
\fl \frac{\partial r_\mu^\nu}{\partial\al_\lb}-\frac{\partial r_\mu^\lb}{\partial\al_\nu}=
\sum_{k,l,i}C_{\mu k}C^\nu_lC^\lb_{i}\left(
\frac{\partial r_{kl}}{\partial a_i}-\frac{\partial r_{ki}}{\partial a_l}\right)=0.
\en
The Jacobian of transformation $\bls{\tau}=\bls{\tau}(\bls{\al})$ is equal to $\sqrt{g(\bls{\al})}>0$, where $g=\det{(\mt{g})}$. The transformation
is then reversible. The reversed transformation is denoted as $\bls{\al}=\bls{\al}(\bls{\tau})$.
The volume element in the space of $\bls{\tau}$ is
\beq\label{voltau}
\rmd\Om (\bls{\tau})= \sqrt{g(\bls{\al})}\rmd\Om (\bls{\al}) .
\eeq

When we assume that the values of $\gm_\mu = \al_\mu - \al_\mu '$ are small, what is consistent with the Gaussian approximation, 
 we have from \eref{tcom}:
\numparts
\bn\label{tau}
  \vth_\mu &=&\tau_\mu -\tau_\mu^\prime  
= r_\mu^\nu \left(\thalf(\bls{\al}+\bls{\al}')\right)(\al_\nu-{\al}'_\nu)=r_\mu^\nu (\bls{\bta})\gm_\nu ,\label{taua} \\
\theta_\mu &=& \thalf (\tau_\mu +\tau^{\prime}_\mu ) = \tau_\mu (\bls{\bta}) , \label{taub}
\en
\endnumparts
where $\tau_\mu '=\tau_\mu (\bls{\al}')$. 

Using the new variables, $\bls{\tau}$, it is easy to solve the eigenvalue equation
\beq\label{evov}
\int\mathcal{I}(\bls{\al}(\bls{\tau}),\bls{\al}(\bls{\tau}'))\chi (\bls{\eps},\bls{\tau}^{\prime})\rmd\Om (\bls{\tau}^{\prime})
=j(\bls{\eps} )\chi (\bls{\eps},\bls{\tau}).
\eeq
for the Gaussian overlap. The wave tensor $\bls{\eps}$ stands for a set of the five quantum numbers,
$\eps_\mu \ (\mu =-2,\ \dots ,\ 2)$, which label eigenfunctions $\chi (\bls{\eps},\bls{\tau})$ and eigenvalues $j(\bls{\eps})$. The normalised eigenfunction reads:
\beq\label{chi}
\chi (\bls{\eps},\bls{\tau})=\frac{1}{(2\pi )^{5/2}}\exp{\left(\rmi\bls{\eps}\cdot\bls{\tau}\right)}.
\eeq
It fulfils the following orthogonality and completeness conditions:
\numparts
\bn
&& \int \chi^{\ast}(\bls{\eps},\bls{\tau})\chi(\bls{\eps}',\bls{\tau})\rmd\Om (\bls{\tau})= \delta_5(\bls{\eps}-\bls{\eps}'), \label{chiort} \\
&& \int \chi^{\ast}(\bls{\eps},\bls{\tau})\chi(\bls{\eps},\bls{\tau}')\rmd\Om (\bls{\eps})= \delta_5(\bls{\tau}-\bls{\tau}'). \label{chicomp}
\en
\endnumparts
The tensor properties of quantum numbers $\eps_\mu$ are clearly seen from \eref{chi}.
The eigenvalue is equal to:
\beq\label{n}
j(\bls{\eps})=(2\pi )^{5/2}\exp{(-\thalf \bls{\eps}\cdot\bls{\eps})}.
\eeq 

\subsection{The orthogonal Hill-Wheeler equation in GOA}\label{HWGOA}
We know that the Hill-Wheeler equation \eref{hw} can be formally reduced to the orthogonal eigenvalue equation (cf \cite{Rin80})
for wave-function $\psi (\bls{\al})$
defined as follows:
\beq\label{expsi}
\psi (\bls{\al})=\int \mathcal{R}(\bls{\al},\bls{\al}')\varphi (\bls{\al}')\rmd\Om (\bls{\al}'),
\eeq
where
\beq\label{capr}
\mathcal{R}(\bls{\al},\bls{\xi})=\int \chi (\bls{\eps},\bls{\tau}(\bls{\al}))j^{1/2}(\bls{\eps})
\chi^{\ast}(\bls{\eps},\bls{\tau}(\bls{\xi}))\rmd\Om (\bls{\eps})
\eeq
is the square root kernel.
The name of $\mathcal{R}$ is justified by the relation
\beq\label{sqrtov}
\int \mathcal{R}(\bls{\al},\bls{\xi})\mathcal{R}(\bls{\xi},\bls{\al}')\rmd\Om (\bls{\tau}(\bls{\xi}))=\mathcal{I}(\bls{\al},\bls{\al}').
\eeq
\Eref{expsi} can be formally reversed and the weight function, $\varphi$ can be expressed in the form:
\beq\label{psi}
\varphi (\bls{\al})=\int \mathcal{N}(\bls{\al},\bls{\xi})\psi (\bls{\xi})\rmd\Om (\bls{\tau}(\bls{\xi})) ,
\eeq
where the narrowing kernel \cite{Gri57}
\beq\label{narrow}
\mathcal{N}(\bls{\al},\bls{\xi})=\int \chi (\bls{\eps},\bls{\tau}(\bls{\al}))j^{-1/2}(\bls{\eps})
\chi^{\ast}(\bls{\eps},\bls{\tau}(\bls{\xi}))\rmd\Om (\bls{\eps})
\eeq 
is introduced, which fulfils the following relation:
\beq\label{revn}
\int \mathcal{R}(\bls{\al},\bls{\xi})\mathcal{N}(\bls{\xi},\bls{\al}')\rmd\Om (\bls{\tau}(\bls{\xi}))
= \delta_5(\bls{\tau}(\bls{\al})-\bls{\tau}(\bls{\al}')).
\eeq 
We define the narrowed overlap kernel 
\bn\label{ortov}
\fl \mathcal{J}(\bls{\xi},\bls{\xi}')=
\int\mathcal{N}^{\ast}(\bls{\al},\bls{\xi})\mathcal{I}(\bls{\al}(\bls{\tau}),\bls{\al}'(\bls{\tau}))\mathcal{N}(\bls{\al}',\bls{\xi}')
\rmd\Om (\bls{\tau}(\bls{\al}))\rmd\Om (\bls{\tau}(\bls{\al}')) \nonumber \\
\fl \phantom{\mathcal{J}(\bls{\xi},\bls{\xi}')}=\delta_5 (\bls{\tau}(\bls{\xi})-\bls{\tau}(\bls{\xi}')) ,
\en
which takes the diagonal form.

To obtain the integral eigenvalue equation for function $\psi$ we substitute \eref{psi} for the weight function, $\varphi$ 
in the Hill-Wheeler equation \eref{hw}, multiply it by $\mathcal{N}(\bls{\al},\bls{\xi})$   and integrate with respect to $\bls{\tau}(\bls{\al})$.
Then we get
\beq\label{narrev}
\int\mathcal{K}(\bls{\xi},\bls{\xi}')\psi (\bls{\xi}')\rmd\Om (\bls{\tau}(\bls{\xi}'))=E\psi (\bls{\xi}) ,
\eeq
where 
\bn\label{nek}
\fl \mathcal{K}(\bls{\xi},\bls{\xi}')=\int\mathcal{N}^{\ast}(\bls{\al},\bls{\xi})
\mathcal{H}(\bls{\al},\bls{\al}')\mathcal{N}(\bls{\al}',\bls{\xi}')
\rmd\Om (\bls{\tau}(\bls{\al}))\rmd\Om (\bls{\tau}(\bls{\al}'))
\en
is the narrowed energy kernel.  

\subsection{The Fourier analysis of the Gaussian energy kernel}\label{FAGEK} 
The problem with equation \eref{narrev}, and kernels \eref{narrow} and \eref{nek} is that eigenvalues $j(\bls{\eps})$ of the Gaussian overlap 
kernel tend to
zero when $\bls{\eps}\cdot\bls{\eps}$ tends to infinity. It means that the integral in \eref{narrow} is divergent and we have the "ultra-violet
catastrophe" for $\mathcal{N}$. To make the integral convergent we can introduce a cut-off for $\bls{\eps}\cdot\bls{\eps}$. Instead, we can use operational methods of the distribution theory and treat integrals being the Fourier transforms of the non-integrable functions as convergent due to
oscillations with the infinite frequencies of the integrand and express them by the Dirac delta function and its derivatives, namely
\beq\label{delder}
\frac{1}{2\pi}\int_{-\infty}^{+\infty}f(k)\rme^{\rmi kx}\rmd k =f(-\rmi\rmd /\rmd x)\delta (x).
\eeq 
\Eref{delder} follows from the formula for the derivatives of the delta function represented in the form of the Fourier integral, namely:
\beq\label{delfourier}
\left(-\rmi\frac{\rmd}{\rmd x}\right)^n\delta (x)=\frac{1}{2\pi}\int_{-\infty}^{+\infty}k^n\rme^{\rmi kx}\rmd k
\eeq
for $n=0,\ 1,\ \dots$.

In order to express the Gaussian energy kernel in terms of the delta function and its derivatives we introduce the Fourier transform, 
$\mathcal{F}$, of $\mathcal{H}$ in the following way:
\bn\label{fourier}
\fl \mathcal{F}(\bls{\eps},\bls{\eps}')= \int \chi^{\ast}(\bls{\eps},\bls{\tau}(\bls{\al}))
\mathcal{H}(\bls{\al},\bls{\al}')\chi (\bls{\eps}',\bls{\tau}(\bls{\al}'))
\rmd\Om (\bls{\tau}(\bls{\al}))\rmd\Om (\bls{\tau}(\bls{\al}')).
\en
Substituting \eref{gaussen}, \eref{taua}, \eref{taub} and \eref{chi} into \eref{fourier} we have
\bn\label{fgauss}
\fl \mathcal{F}(\bls{\eps},\bls{\eps}')=\frac{1}{(2\pi )^5}\int [v(\bls{\bta})-\thalf a^{\kappa\lb}(\bls{\bta})
\vth_\kappa (\bls{\bta},\bls{\gm})\vth_\lb (\bls{\bta},\bls{\gm})] \nonumber \\
\fl \times\exp{(-\rmi\bls{\sigma}\cdot\bls{\theta}(\bls{\bta}))}\exp{(-\thalf\bls{\vth}(\bls{\bta},\bls{\gm})
\cdot\bls{\vth}(\bls{\bta},\bls{\gm})-\rmi\bls{\veps}\cdot\bls{\vth}(\bls{\bta},\bls{\gm}))}
\rmd\Omega (\bls{\theta}(\bls{\bta}))\rmd\Omega (\bls{\vth}(\bls{\bta},\bls{\gm})),
\en
where
\beq\label{a}
a_{\kappa\lb}(\bls{\bta})=h^{\mu\nu}(\bls{\bta})n_{\mu\kappa}(\bls{\bta})n_{\lb\nu} (\bls{\bta})
\eeq
and $\bls{\veps}=\thalf(\bls{\eps}+\bls{\eps}')$, $\bls{\sigma}=\bls{\eps}-\bls{\eps}'$, and 
$n_{\mu\nu}=(1/{\sqrt{g}})(\partial\sqrt{g}/\partial r^{\mu\nu})$ is the matrix reversed to $\mt{r}$, i.e.
$n_{\nu\lb}r^{\lb\mu}=\delta^\mu_\nu$. The Gaussian integrals with respect to $\bls{\vth}$ can be easily calculated and we obtain 
\bn\label{fourint}
\fl \mathcal{F}(\bls{\eps},\bls{\eps}')= \frac{j(\bls{\veps})}{(2\pi )^5}\int [v(\bls{\bta})
+\thalf(\veps^\kappa\veps_\lb - \delta^\kappa_\lb )a_{\kappa}^\lb(\bls{\bta})] 
\exp{(-\rmi\bls{\sigma}\cdot\bls{\theta}(\bls{\bta}))}\rmd\Om (\bls{\theta}(\bls{\bta})).
\en

The narrowed energy kernel \eref{nek} expressed in terms of the Fourier transform \eref{fourier} reads:
\bn\label{kf}
\fl \mathcal{K}(\bls{\xi},\bls{\xi}')=\int j^{-1/2}(\bls{\eps})j^{-1/2}(\bls{\eps}')\chi (\bls{\eps},\bls{\xi})
\mathcal{F}(\bls{\eps},\bls{\eps}')\chi^{\ast}(\bls{\eps}',\bls{\xi}')\rmd\Om (\bls{\eps})\rmd\Om (\bls{\eps}').
\en
Inserting \eref{chi}, \eref{n} and \eref{fourint} into \eref{kf} we have 
\bn\label{kint}
\fl \mathcal{K}(\bls{\xi},\bls{\xi}')=\frac{1}{(2\pi )^{10}}\int\exp{(\case{1}{8}\bls{\sigma}\cdot\bls{\sigma})}
\exp{(\rmi\bls{\sigma}\cdot (\bls{\theta}(\bls{\eta})-\bls{\theta}(\bls{\bta}))}\rmd\Om (\bls{\sigma}) \nonumber \\
\fl  \times[v(\bls{\bta})
+\thalf(\veps^\kappa\veps_\lb - \delta^\kappa_\lb )a_\kappa^\lb (\bls{\bta})]
\exp{(\rmi\bls{\veps}\cdot\bls{\vth}(\bls{\eta},\bls{\zeta}))}\rmd\Om (\bls{\veps})\rmd\Om (\bls{\theta}(\bls{\bta})),
\en
where $\bls{\eta}=\thalf (\bls{\xi}+\bls{\xi}')$ and $\bls{\zeta}=\bls{\xi}-\bls{\xi}'$.
The integrals with respect to $\bls{\veps}$ and $\bls{\sigma}$ in \eref{kint} are the fivefold integrals of type \eref{delder} and thus can be expressed  
in terms of the partial derivatives of the five-dimensional Dirac $\delta$
function in the following way:
\bn
\fl \frac{1}{(2\pi )^5}\int[v(\bls{\bta})
+\thalf(\veps^\kappa\veps_\lb - \delta^\kappa_\lb )a^\lb_\kappa(\bls{\bta})]
\exp{(\rmi\bls{\veps}\cdot\bls{\vth}(\bls{\eta},\bls{\zeta}))}
\rmd\Om (\bls{\veps}) \nonumber \\
\fl =\left[v(\bls{\bta})+\thalf\left(\left(-\rmi\frac{\partial}{\partial\vth_\kappa (\bls{\eta},\bls{\zeta})}\right)
\left(-\rmi\frac{\partial}{\partial\vth^\lb (\bls{\eta},\bls{\zeta})}\right)
 - \delta^\kappa_\lb \right)a^\lb_\kappa(\bls{\bta})\right]\delta_5(\bls{\vth}(\bls{\eta},\bls{\zeta})),\label{intveps} \\
\frac{1}{(2\pi )^5}\int\exp{(\case{1}{8}\bls{\sigma}\cdot\bls{\sigma})}\exp{(\rmi\bls{\sigma}\cdot (\bls{\theta}(\bls{\eta})-\bls{\theta}(\bls{\bta}))}
\rmd\Om (\bls{\sigma}) \nonumber \\
=\exp{\left(\case{1}{8}\left(-\rmi\frac{\partial}{\partial\theta_\mu (\bls{\eta})}\right)
\left(-\rmi\frac{\partial}{\partial\theta^\mu (\bls{\eta})}\right)\right)}
\delta_5(\bls{\theta}(\bls{\eta})-\bls{\theta}(\bls{\bta})).\label{intsigma}
\en
After inserting \eref{intveps} and \eref{intsigma} into \eref{kint} the integration with respect to $\bls{\theta}(\bls{\bta})$ is easily performed and
finally the narrowed energy kernel takes the form:
\bn\label{kdelta}
\fl \mathcal{K}(\bls{\xi},\bls{\xi}') 
 =\exp{(-\case{1}{8}\Delta (\bls{\theta}(\bls{\eta})))}\nonumber \\ 
\fl \times\left[v(\bls{\eta})
-\thalf h^{\mu\nu}(\bls{\eta})f_{\mu\nu}(\bls{\eta})
 -\thalf a_\kappa^\lb(\bls{\eta})\frac{\partial}{\partial\vth^\kappa (\bls{\eta},\bls{\zeta})}
\frac{\partial}{\partial\vth^\lb (\bls{\eta},\bls{\zeta})}
\right]\delta_5(\bls{\vth}(\bls{\eta},\bls{\zeta})) ,
\en
where $\mt{f}=\mt{g}^{-1}=\mt{n}\cdot\mt{n}$ is the matrix reversed to $\mt{g}$, and
\beq\label{laplt} 
\Delta (\bls{\tau}) = \frac{\partial}{\partial\tau_\mu }\frac{\partial}{\partial\tau^\mu }.
\eeq
The kernel is thus the distribution dependent on the delta function of $\bls{\vth}(\bls{\xi},\bls{\xi}')= \bls{\tau}(\bls{\xi}) -\bls{\tau}(\bls{\xi}')$ 
and its partial derivatives of the second order. We rewrite it in a shorter form:
\bn\label{kdelshort}
\fl \mathcal{K}(\bls{\xi},\bls{\xi}') =
\left[\bar{v}'(\bls{\eta})
 -\thalf \bar{a}_\kappa^\lb(\bls{\eta})\frac{\partial}{\partial\vth^\kappa (\bls{\eta},\bls{\zeta})}
\frac{\partial}{\partial\vth^\lb (\bls{\eta},\bls{\zeta})}
\right]\delta_5(\bls{\vth}(\bls{\eta},\bls{\zeta})),
\en 
where
\numparts
\bn\label{avbar}
 \bar{a}_{\mu\nu}(\bls{\eta})=\exp{(-\case{1}{8}\Delta (\bls{\theta}(\bls{\eta})))}a_{\mu\nu}(\bls{\eta}), \label{abar}\\
\fl \bar{v}'(\bls{\eta})=\exp{(-\case{1}{8}\Delta (\bls{\theta}(\bls{\eta})))}\left[v(\bls{\eta})
-\thalf h^{\mu\nu}(\bls{\eta})f_{\mu\nu}(\bls{\eta})\right]=\exp{(-\case{1}{8}\Delta (\bls{\theta}(\bls{\eta})))}v'(\bls{\eta}). 
 \label{vpbar}
\en
\endnumparts

\section{Differential eigenvalue equations}\label{DE}
\subsection{The differential Hill-Wheeler equation}\label{DHW}
The integral Hill-Wheeler equation \eref{narrev} is easily converted into a differential equation using the kernel in the form of \eref{kdelta}. Indeed, inserting \eref{kdelta} into \eref{narrev} we have: 
\bn\label{hwdelta}
\fl \int \psi (\bls{\xi}')\left[\bar{v}'\left(\thalf (\bls{\xi}+\bls{\xi}')\right)
 -\thalf \bar{a}_\kappa^\lb(\thalf (\bls{\xi}+\bls{\xi}'))\frac{\partial}{\partial\tau^\kappa (\bls{\xi}')}
\frac{\partial}{\partial\tau^\lb (\bls{\xi}')}
\right]\delta_5(\bls{\tau}(\bls{\xi}) -\bls{\tau}(\bls{\xi}'))
\rmd\Om (\bls{\tau}(\bls{\xi}')) \nonumber \\
=E\psi (\bls{\xi}).
\en
Integrating by parts we obtain from \eref{hwdelta} the following differential
eigenvalue equation:
\beq\label{diftau}
\left[-\half\frac{\partial}{\partial\tau_\mu (\bls{\xi})}\bar{a}_{\mu\nu}(\bls{\xi})\frac{\partial}{\partial\tau_\nu (\bls{\xi})}
+\bar{V}(\bls{\xi}(\bls{\tau}))\right]{\psi}(\bls{\xi}) = E{\psi}(\bls{\xi}),
\eeq
where
\beq
\bar{V}(\bls{\xi})=\bar{v}'(\bls{\xi})
 -\frac{1}{8}\frac{\partial^2\bar{a}_{\mu\nu}(\bls{\xi})}{\partial\tau_\mu (\bls{\xi})\partial\tau_\nu (\bls{\xi})}.  \label{vbar}
\eeq

Coming back to the original variables $\bls{\xi}$ in \eref{diftau} we transform the derivatives with respect to $\tau_\mu (\bls{\xi})$ 
into the derivatives with respect to $\xi_\mu$ and we obtain the eigenvalue equation
\beq\label{evbar}
\bar{H}(\bls{\xi}){\psi}(\bls{\xi})= E\psi (\bls{\xi})
\eeq
for the Bohr Hamiltonian
\beq\label{hbar}
\bar{H}(\bls{\xi})=-\frac{1}{2\sqrt{g(\bls{\xi})}}\frac{\partial}{\partial\xi_\mu}
\sqrt{g(\bls{\xi})}\bar{A}_{\mu\nu}(\bls{\xi})\frac{\partial}{\partial\xi_\nu}+\bar{V}(\bls{\xi})
\eeq
with the inverse inertial bitensor 
\beq\label{invinertbar}
\bar{A}_{\mu\nu}(\bls{\xi})=n_{\mu\kappa}(\bls{\xi})n_{\nu\lb}(\bls{\xi})\bar{a}^{\kappa\lb}(\bls{\xi})
\eeq
and the potential
\bn\label{potbar}
\fl \bar{V}(\bls{\xi})=\exp{(-\case{1}{8}\Delta (\bls{\xi}))}
v'(\bls{\xi}) 
 -\frac{1}{8}\frac{1}{\sqrt{g(\bls{\xi})}}\frac{\partial}{\partial\xi_\mu}\left(\sqrt{g(\bls{\xi})}n_{\mu\kappa}(\bls{\xi})n_{\nu\lb}(\bls{\xi})
\frac{\partial\bar{a}^{\kappa\lb}(\bls{\xi})}{\partial\xi_\nu}\right) .
\en
The Laplacian \eref{laplt} takes the form:
\beq\label{laple}
\Delta (\bls{\xi}) = \frac{1}{\sqrt{g(\bls{\xi})}}\frac{\partial}{\partial\xi_\mu}\sqrt{g(\bls{\xi})}f_{\mu\nu}(\bls{\xi})
\frac{\partial}{\partial\xi_\nu}.
\eeq
The relation 
\beq\label{difsqq}
\frac{\partial}{\partial\xi_\mu}\left(\sqrt{g(\bls{\xi})}n_{\mu\nu}(\bls{\xi})\right)=0,
\eeq
which comes from \eref{wl}, has been used to obtain \eref{hbar}, \eref{potbar} and \eref{laple}.

\Eref{evbar} is equivalent to the Hill-Wheeler equation \eref{narrev} with
the Gaussian plus quadratic approximation \eref{gaussover} and \eref{gaussen} for the kernels. No other approximation is made. However, one can say that the equivalence is formal only because the quantities \eref{invinertbar} and \eref{potbar} can be determined merely in an approximation 
when expanding the exponential operator. For instance, we expand \eref{abar}:
\bn\label{exa}
\fl \bar{a}_{\mu\nu}(\bls{\xi})={a}_{\mu\nu}(\bls{\xi})-\frac{1}{8}\Delta (\bls{\xi}){a}_{\mu\nu}(\bls{\xi})
+\half\left(\frac{1}{8}\right)^2\Delta (\bls{\xi})\Delta (\bls{\xi}){a}_{\mu\nu}(\bls{\xi})+\dots
\en
However, we have no guarantee whether such an approximation procedure is convergent.

\subsection{The collective Hamiltonian}\label{CH}
Another method of deriving the Bohr Hamiltonian, which passes by the integral equation \eref{narrev} and starts
directly from the variational principle, is more often used (cf \cite{Rin80,Goz85b,Rei87}). We
refer to it as the collective approach to the GCM.
It consists in representing  the energy mean value in the form of the expectation value of a local
differential operator, say $H$, Hermitian with weight $w$ in the space of the generator coordinates, namely:
\bn\label{collmod}
 \langle\Psi [\varphi ]|\hat{H}|\Psi [\varphi ]\rangle &=& 
\int\varphi^\ast (\bls{\al})\mathcal{H}(\bls{\al},\bls{\al}')\varphi (\bls{\al}')\rmd\Omega (\bls{\al})\rmd\Omega (\bls{\al}') \nonumber \\
&=&\int\psi^\ast (\bls{\al})H(\bls{\al})\psi (\bls{\al})w(\bls{\al})\rmd\Omega (\bls{\al}).
\en
In the GOA the square root kernel \eref{capr} is equal to
\beq\label{rgoa}
\mathcal{R}(\bls{\al},\bls{\xi})=\left(\frac{2}{\pi}\right)^{5/4}
\exp{\left(-\left(\bls{\tau}(\bls{\al})-\bls{\tau}(\bls{\xi})\right)\cdot \left(\bls{\tau}(\bls{\al})-\bls{\tau}(\bls{\xi})\right)\right)}
\eeq
and then, according to \eref{sqrtov} and \eref{gaussen} the energy kernel can be presented in the following form
\bn\label{engoa}
\fl \mathcal{H}(\bls{\al},\bls{\al}')=\int\mathcal{R}(\bls{\al},\bls{\xi})[v(\bls{\bta})-\half h^{\mu\nu}(\bls{\bta})\gm_\mu\gm_\nu ]
\mathcal{R}(\bls{\xi},\bls{\al}')\sqrt{g(\bls{\xi})}\rmd\Omega (\bls{\xi}) \nonumber \\
\fl =\int\mathcal{R}(\bls{\al},\bls{\xi})[v(\bls{\bta})-\half a^{\mu\nu}(\bls{\bta})
\vth_\mu (\bls{\al},\bls{\al}')\vartheta_\nu (\bls{\al},\bls{\al}')]
\mathcal{R}(\bls{\xi},\bls{\al}')\rmd\Omega (\bls{\tau}(\bls{\xi})).
\en

An additional approximation should be made to achieve the form \eref{collmod} for the energy functional. Namely, we assume that functions $v(\bls{\bta})$ and $a^{\mu\nu}(\bls{\bta})$ do not change much
within the range of $\bls{\bta}=\thalf(\bls{\al}+\bls{\al}')$ in the vicinity of $\bls{\xi}$ where both Gaussian functions, $\mathcal{R}(\bls{\al},\bls{\xi})$ 
and $\mathcal{R}(\bls{\xi},\bls{\al}')$ take still big values. In other words we assume that $v(\bls{\bta})$ and $a^{\mu\nu}(\bls{\bta})$ are slowly varying
functions in the vicinity of $\bls{\xi}$, and we can expand them in the power series around $\bls{\xi}$:
\numparts
\bn
\fl v(\bls{\bta})\approx v(\bls{\xi})+\frac{\partial v(\bls{\xi})}{\partial\xi_\mu}(\bta_\mu-\xi_\mu)
+\thalf\frac{\partial^2 v(\bls{\xi})}{\partial\xi_\mu\partial\xi_\nu}(\bta_\mu-\xi_\mu)(\bta_\nu-\xi_\nu) +\dots \label{slowv} \\ 
\fl a^{\mu\nu}(\bls{\bta})\approx a^{\mu\nu}(\bls{\xi})+\frac{\partial a^{\mu\nu}(\bls{\xi})}{\partial\xi_\kappa}(\bta_\kappa-\xi_\kappa)
+\thalf\frac{\partial^2 a^{\mu\nu}(\bls{\xi})}{\partial\xi_\kappa\partial\xi_\lb}(\bta_\kappa-\xi_\kappa)(\bta_\lb-\xi_\lb) +\dots \label{slowa}
\en
\endnumparts 
in \eref{engoa}. Here, we will exploit the lowest approximation and take the zero-order terms only. Since the Gaussian functions with monomial 
factors in front are equal to the derivatives of the Gaussian functions themselves, the $\bls{\al}$- and $\bls{\al}'$-dependence of the integrand in \eref{engoa}
appear only in functions $\mathcal{R}(\bls{\al},\bls{\xi})$ and $\mathcal{R}(\bls{\xi},\bls{\al}')$ and their derivatives with respect to $\xi_\mu$.
 Then, the integration with respect $\bls{\al}$ and $\bls{\al}'$ in \eref{collmod} can be performed giving
the collective wave-functions $\psi^\ast (\bls{\xi})$ and $\psi (\bls{\xi})$ of \eref{expsi}. Finally, we can obtain the energy mean value in the 
form \eref{collmod} performing integrations by parts. The collective Hamiltonian in the space of coordinates $\bls{\tau}(\bls{\xi})$ reads
\beq\label{hcoolt}
H =-\half\frac{\partial}{\partial\tau_\mu (\bls{\xi})}{a}_{\mu\nu}(\bls{\xi})\frac{\partial}{\partial\tau_\nu (\bls{\xi})}
+{V}(\bls{\xi})
\eeq
with $a_{\mu\nu}$ given by \eref{a} and
\bn\label{pottau}
V(\bls{\xi})&=&
v(\bls{\xi})
-\thalf h^{\mu\nu}(\bls{\xi})f_{\mu\nu}(\bls{\xi}) 
 -\frac{1}{8}\frac{\partial^2a_{\mu\nu}(\bls{\xi})}{\partial\tau_\mu (\bls{\xi})\partial\tau_\nu (\bls{\xi})} \nonumber \\
 &=&v'(\bls{\xi})
 -\frac{1}{8}\frac{\partial^2a_{\mu\nu}(\bls{\xi})}{\partial\tau_\mu (\bls{\xi})\partial\tau_\nu (\bls{\xi})}.
\en
Coming back to variables $\bls{\xi}$ we end up with the collective Bohr Hamiltonian
\beq\label{collham}
H=-\frac{1}{2\sqrt{g(\bls{\xi})}}\frac{\partial}{\partial\xi_\mu}\sqrt{g(\bls{\xi})}{A}_{\mu\nu}(\bls{\xi})\frac{\partial}{\partial\xi_\nu}
+{V}(\bls{\xi}),
\eeq
Hermitian with weight $w(\bls{\xi})= \sqrt{g(\bls{\xi})}$, where
\beq\label{invinertcoll}
A_{\mu\nu}(\bls{\xi})=f_{\mu\kappa}(\bls{\xi})f_{\nu\lb}(\bls{\xi})h^{\kappa\lb}(\bls{\xi})
\eeq
and
\bn\label{potcoll}
 V(\bls{\xi})&=&v(\bls{\xi})
-\thalf h^{\mu\nu}(\bls{\xi})f_{\mu\nu}(\bls{\xi}) \nonumber \\ 
&& -\frac{1}{8}\frac{1}{\sqrt{g(\bls{\xi})}}\frac{\partial}{\partial\xi_\mu}\left(\sqrt{g(\bls{\xi})}
n_{\mu\kappa}(\bls{\xi})n_{\nu\lb}(\bls{\xi})\frac{\partial a^{\kappa\lb}(\bls{\xi})}{\partial\xi_\nu}\right) .
\en

We see that the Bohr Hamiltonians $\bar{H}$ and $H$ have the same form which is identical with that discussed in \cite{Pro09} for
the quantum collective models. The only difference between them is that they are determined by different inverse inertial bitensors
and different collective potentials, $\bar{A}_{\mu\nu}$ and $A_{\mu\nu}$, and $\bar{V}$ and $V$, respectively. For $\exp{(-\case{1}{8}\Delta )}\approx 1$ the quantities with bar transform into the corresponding quantities without
bar. Such an approximation is good when $a_{\mu\nu}(\bls{\xi})$ and $v(\bls{\xi})$ are slowly varying functions
of their argument. It is in accordance with the assumption made when deriving Hamiltonian $H$ of \eref{collham}.  
 
\subsection{The zero-point energy}\label{ZPE}
In the original approach to the Bohr Hamiltonian, there is an ambiguity related to the correspondence principle. The Podolsky-Pauli quantization prescription
tells us how to quantize the kinetic energy; however, in general, the kinetic energy operator is given up to an additive arbitrary function of the deformation
\cite{Hof72}. In consequence, the potential energy which enters the Bohr Hamiltonian operator needs not to be identical with the classical one,
as has been often assumed in the past (cf \cite{Pro09}). In the present study it would correspond to the collective potential 
$V(d)= v(d)=\langle\phi (d)|\hat{H}|\phi (d)\rangle$ equal to the static ground-state energy term . On the other hand, it is obvious that dynamical 
correlations in the ground state of quantum systems should appear. Various effects of the ground-state correlations have been investigated for a long time
(see, e.g., a review \cite{Rei94}). In the theory of the collective excitations, the difference between the static energy and the collective potential is 
interpreted as the zero-point energy associated with a given collective mode. The GCM gives a definite result for the zero-point energy correction to the static
ground-state energy. The zero-point energies associated with the quadrupole vibrational modes have been already estimated in the frame of the GOA, e.g., 
\cite{Rei75,Gir79,Bon90,Bon91,Hag03}. Those associated with the rotational modes were investigated rather by means of the angular-momentum projection technique
\cite{Bon91,Hag03}.  

In the present five-dimensional GCM with the set of generator coordinates forming the quadrupole tensor, all the quadrupole modes, both vibrational and
rotational are treated on an equal footing. The collective potential of \eref{potcoll} contains the zero-point energy correction, $v(\bls{\xi})-V(\bls{\xi})$,
associated with all the five quadrupole modes. We see from \eref{potcoll} that the correction is composed of two terms. The first, $\thalf\mt{f}\cdot\mt{h}$, 
is connected with the quadratic term in the energy kernel expansion \eref{gaussen} and manifests the dynamical effects. The second one appears only when the
coefficients in the Gaussian expansions \eref{gaussover} and \eref{gaussen}, $g_{\mu\nu}(\bls{\bta})$ and $h_{\mu\nu}(\bls{\bta})$, depend essentially on $\bls{\bta}$ (the case of the local Gaussian approximation). The additional zero-point energy corrections come from expanding the exponential Laplace operator in \eref{avbar} and \eref{vpbar} as discussed in \sref{DHW}. Then, additional terms appear in both, the inverse inertial bitensor, $\bar{A}_{\mu\nu}$, and 
the potential, $\bar{V}$, in the Bohr Hamiltonian $\bar{H}$ of \eref{hbar}. It seems that corrections of that type to the zero-point energy have been taken into account in \cite{Bon90} for the one-dimensional Bohr Hamiltonian. In our case of the five-dimensional quadrupole modes a counterpart of the collective potential with similar corrections would have perhaps the form:
\bn\label{vbon}
\fl  \bar{V}(\bls{\xi})=(1-\case{1}{8}\Delta (\bls{\xi}))
v'(\bls{\xi}) 
 -\frac{1}{8}\frac{1}{\sqrt{g(\bls{\xi})}}\frac{\partial}{\partial\xi_\mu}\left(\sqrt{g(\bls{\xi})}n_{\mu\kappa}(\bls{\xi})n_{\nu\lb}(\bls{\xi})
\frac{\partial a^{\kappa\lb}(\bls{\xi})}{\partial\xi_\nu} \right) .   
\en

The standard procedure of deriving the collective Hamiltonian, shown in \cite{Rin80,Goz85b,Rei87}, consists in expanding 
quotient $\mathcal{H}(\bls{\al},\bls{\al}')/\mathcal{I}(\bls{\al},\bls{\al}')$ of \eref{gaussen} (the expression inside the square brackets in \eref{engoa}) as
a function of $\bls{\al}$ and $\bls{\al}'$ in the power series around an arbitrary point $\bls{\xi}$ up to the second order. The actual meaning of the approximation is not very transparent. Certainly, it constitutes a stronger approximation than the usual quadratic approximation of \eref{gaussen} in the GOA. It seems that a counterpart
of such an approximation in the present approach is expanding the static potential
\eref{slowv} up to the second order. Then, the resulting collective potential would have the form: 
\bn\label{vring}
  \bar{V}(\bls{\xi})&=&(1-\case{1}{8}\Delta (\bls{\xi}))
v(\bls{\xi})-\thalf h^{\mu\nu}(\bls{\xi})f_{\mu\nu}(\bls{\xi}) \nonumber \\ 
 &&-\frac{1}{8}\frac{1}{\sqrt{g(\bls{\xi})}}\frac{\partial}{\partial\xi_\mu}\left(\sqrt{g(\bls{\xi})}n_{\mu\kappa}(\bls{\xi})n_{\nu\lb}(\bls{\xi})
\frac{\partial a^{\kappa\lb}(\bls{\xi})}{\partial\xi_\nu} \right) .   
\en
 
\subsection{Matrix elements of observables}\label{O}
Solving the Bohr equation \eref{evbar} (or that with Hamiltonian \eref{collham}) provides us with the energies, $E_N$, and the collective wave-functions,  $\psi_N$, of the collective states 
\beq\label{nstate}
|\Psi_N\rangle = \int \mathcal{N}(\bls{\al},\bls{\xi}){\psi}_N(\bls{\xi})|\Phi (\bls{\al})\rangle \sqrt{g(\bls{\xi})}
\rmd\Om (\bls{\xi})\sqrt{g(\bls{\al})}\rmd\Om (\bls{\al}),
\eeq
where $N$ stands for a set of quantum numbers of the state. 
The states form an orthonormal set, i.e., $\langle\Psi_N | \Psi_{N'}\rangle = \delta_{NN'}$ when 
\beq\label{psinorm}
\int \psi^\ast_N (\bls{\xi})\psi_{N'}(\bls{\xi})\sqrt{g(\bls{\xi})}\rmd\Omega (\bls{\xi})= \delta_{NN'} .
\eeq

To know other characteristics of the collective states we should calculate the matrix 
elements of the other observables. In the GOA we approximate the matrix element of observable $\hat{Q}$ within states $|\Phi (\bls{\al})\rangle$
in the way similar to \eref{gaussen}:
\beq\label{gaussq}
\langle\Phi (\bls{\al})|\hat{Q}|\Phi (\bls{\al}')\rangle =q(\bls{\bta})+\rmi q^\mu (\bls{\bta})\gm_\mu -\half q^{\mu\nu}(\bls{\bta})\gm_\mu\gm_\nu + \dots
\eeq
Proceeding further in the way similar to that of sections \ref{FAGEK} and \ref{DHW} for $\hat{H}$ we obtain
\bn\label{matqbar}
 \langle\Psi_N|\hat{Q}|\Psi_{N'}\rangle & =& \int {\psi}_N^\ast (\bls{\xi})\left[
-\frac{1}{2}\frac{\partial}{\partial\xi_\mu}\sqrt{g(\bls{\xi})}\bar{Q}_{\mu\nu}(\bls{\xi})\frac{\partial}{\partial\xi_\nu}\right. \nonumber \\
&& -\frac{\rmi}{2}\left(\frac{\partial}{\partial\xi_\mu}\sqrt{g(\bls{\xi})}\bar{Q}_{\mu}(\bls{\xi}) 
+\sqrt{g(\bls{\xi})}\bar{Q}_{\mu}(\bls{\xi}) \frac{\partial}{\partial\xi_\mu}\right) \nonumber \\
&& \left.+\sqrt{g(\bls{\xi})}\bar{Q}(\bls{\xi})\right]{\psi}_{N'}(\bls{\xi}) \rmd\Omega (\bls{\xi}),
\en
where $\bar{Q}_{\mu\nu}$ and $\bar{Q}$ are given by \eref{invinertbar} and \eref{potbar} when $h^{\al\bta}$ and $v$ are replaced 
with $q^{\al\bta}$ and $q$, respectively,
whereas
\beq\label{qm}
\bar{Q}_{\mu}(\bls{\xi})=n_{\mu\kappa}(\bls{\xi})\exp{(-\case{1}{8}\Delta (\bls{\xi}))}\left[
q^{\al}(\bls{\xi})n^\kappa_\al (\bls{\xi})\right], 
\eeq
and $\psi_N$ are eigenfunctions of Hamiltonian $\bar{H}$.

When $q$, $n_\mu^\nu q_\nu$ and $n_\mu^\kappa q_{\kappa\lb}n^\lb_\nu$ are slowly varying functions of $\bls{\xi}$ the exponential Laplacian operator 
can be replaced by the unity and we have the expression similar in form with \eref{matqbar}(cf \cite{Ner86}), namely:
\bn\label{matq}
 \langle\Psi_N|\hat{Q}|\Psi_{N'}\rangle & =& \int {\psi}_N^\ast (\bls{\xi})\left[
-\frac{1}{2}\frac{\partial}{\partial\xi_\mu}\sqrt{g(\bls{\xi})}{Q}_{\mu\nu}(\bls{\xi})\frac{\partial}{\partial\xi_\nu}\right. \nonumber \\
&& -\frac{\rmi}{2}\left(\frac{\partial}{\partial\xi_\mu}\sqrt{g(\bls{\xi})}{Q}_{\mu}(\bls{\xi}) 
+\sqrt{g(\bls{\xi})}{Q}_{\mu}(\bls{\xi}) \frac{\partial}{\partial\xi_\mu}\right) \nonumber \\
&& \left.+\sqrt{g(\bls{\xi})}{Q}(\bls{\xi})\right]{\psi}_{N'}(\bls{\xi}) \rmd\Omega (\bls{\xi}),
\en
where $Q_{\mu\nu}$, $Q_\mu$ and $Q$  are given by the corresponding formulae for $\bar{Q}_{\mu\nu}$, $\bar{Q}_\mu$ and $\bar{Q}$ with the exponential operator put equal to unity, and $\psi_N$ are this time eigenfunctions of Hamiltonian $H$.

\section{The deformation dependence of the quadrupole collective Hamiltonian}\label{CK}
The original object in the present approach to the description of the quadrupole collective states in even-even nuclei
is a set of the many-body intrinsic states $|\phi (d)\rangle$ parametrised by the two quadrupole deformations $d=(d_0,d_2)$.
Knowing $|\phi (d)\rangle$ we can express the overlap and energy kernels, the two basic quantities in the Hill-Wheeler equation,
in terms of overlaps and matrix elements of $\hat{H}$ and $\hat{J}_u$ within states $|\phi (d)\rangle$  themselves and their derivatives \eref{deriv}.
All these overlaps and matrix elements are functions of the deformation. Having the kernels as functions of the deformation we calculate the collective potential and inertial functions which determine the Bohr Hamiltonian.
  
\subsection{The overlap and energy kernels as functions of deformation}\label{GH}
The overlap kernel $\mathcal{I}(\bls{\al},\bls{\al}')$ is real, symmetric and normalised to unity for the all values of variables $\al_\mu$, namely
\numparts
\beq
\mathcal{I}(\bls{\al},\bls{\al})=\langle\Phi (\bls{\al})|\Phi (\bls{\al})\rangle = 1.\label{norm0}
\eeq
Differentiating twice \eref{norm0} we have the following conditions:
\bn
&&\langle\Phi (\bls{\al})|\Phi^\mu (\bls{\al})\rangle = 0,\label{norm1}\\
&&\langle\Phi (\bls{\al})|\Phi^{\mu\nu} (\bls{\al})\rangle +\langle\Phi^\mu (\bls{\al})|\Phi^\nu (\bls{\al})\rangle = 0,\label{norm2}
\en
\endnumparts
where
\beq\label{deriva}
 |\Phi^\mu (\bls{\al})\rangle =\frac{\partial}{\partial\al_\mu}|\Phi (\bls{\al})\rangle ,\quad
|\Phi^{\mu\nu} (\bls{\al})\rangle =\frac{\partial^2}{\partial\al_\mu\partial\al_\nu}|\Phi (\bls{\al})\rangle
\eeq
are the states created by differentiation with respect to the components of tensor $\bls{\al}$.

The GOA consists technically in expanding the logarithm of overlap kernel in powers of $\bls{\gm}=\bls{\al}-\bls{\al}'$ up to the second order in $\bls{\gm}$.
The coefficients of expansion are the partial derivatives of $\ln{\mathcal{I}(\bls{\bta}+\thalf\bls{\gm},\bls{\bta}-\thalf\bls{\gm})}$ at $\bls{\gm}=0$.
They are:
\numparts
\bn\label{firstder}
&& \left.\frac{\partial\ln{\mathcal{I}(\bls{\bta}+\thalf\bls{\gm},\bls{\bta}-\thalf\bls{\gm})}}
{\partial\gm_\mu}\right|_{\gm_{-2}=\dots =\gm_2 =0}
\equiv\frac{\partial\ln{\mathcal{I}(\bls{\bta},\bls{\bta})}}{\partial\gm_\mu} \nonumber \\
&&=\frac{\langle\Phi^\mu (\bls{\bta})|\Phi (\bls{\bta})\rangle -\langle\Phi (\bls{\bta})|\Phi^\mu (\bls{\bta})\rangle}
{2\langle\Phi (\bls{\bta})|\Phi (\bls{\bta})\rangle} =0
\en
in virtue of \eref{norm0} and \eref{norm1}, and in accordance with \eref{lnover}.
\bn\label{secder}
\fl \frac{\partial^2\ln{\mathcal{I}(\bls{\bta},\bls{\bta})}}{\partial\gm_\mu\partial\gm_\nu}
=-\left(\mathcal{I}(\bls{\bta},\bls{\bta})\right)^{-2}\frac{\partial\mathcal{I}(\bls{\bta},\bls{\bta})}{\partial\gm_\mu}
\frac{\partial\mathcal{I}(\bls{\bta},\bls{\bta})}{\partial\gm_\nu}
+\left(\mathcal{I}(\bls{\bta},\bls{\bta})\right)^{-1}\frac{\partial^2\mathcal{I}(\bls{\bta},\bls{\bta})}{\partial\gm_\mu\partial\gm_\nu} \nonumber \\
=\frac{\partial^2\mathcal{I}(\bls{\bta},\bls{\bta})}{\partial\gm_\mu\partial\gm_\nu}
\en
\endnumparts
according to \eref{norm0} and \eref{firstder}. Hence, the metric tensor defined by \eref{lnover} is
\beq\label{gdef}
g^{\mu\nu}(\bls{\bta})=-\frac{\partial^2\mathcal{I}(\bls{\bta},\bls{\bta})}
{\partial\gm_\mu\partial\gm_\nu}.
\eeq
Performing the second order derivative and using additionally \eref{norm2} we have finally
\beq\label{gdeffin}
g^{\mu\nu}(\bls{\al})= \langle\Phi^\mu (\bls{\al})|\Phi^\nu (\bls{\al})\rangle .
\eeq 
State $|\Phi(\bls{\al})\rangle$ is of the form of \eref{rotstate}. Thus, using \eref{dalpha} we replace the derivatives with respect to $\al_\mu$ in
\eref{gdeffin} by the derivatives with respect to the Euler
angles $\omega$ and deformations $d$. The $\omega$-dependence of $|\Phi (\bls{\al})\rangle$ is inherent in $\hat{R}(\omega )$.
Differentiating the rotation operator with respect to the Euler angles we obtain the formula
\beq\label{domega}
L_u(\omega)\hat{R}(\omega ) = -\hat{R}(\omega )\hat{J}_u\quad \mathrm{for}\; u=x,y,z.
\eeq
Finally, $g^{\mu\nu}(\bls{\al})$ is obtained in the following form (cf \eref{mat}):
\beq\label{glab}
g^{\mu\nu}(\bls{\al} )=\sum_{a,b}D^\mu_a (\omega )D^\nu_b (\omega )g_{ab}(d),
\eeq 
where $a,\ b=0,\ 2,\ x,\ y,\ z$.
The intrinsic Cartesian components are:
\numparts
\bn\label{gint}
&&g_{kl}(d)=\langle\phi_k (d)|\phi_l(d)\rangle \quad \mathrm{for}\; k,l=0,2, \label{ginta}\\
&&g_{uu}(d)=\frac{1}{4d_u^2}\langle\phi (d)|\hat{J}^2_u|\phi (d)\rangle\quad \mathrm{for}\; u=x,y,z \label{gintb}\\
&&g_{ku}(d)=g_{uu'}(d)=0 \quad \mathrm{for}\; k=0,2,\; u,u'=x,y,z,\; u\ne u', \label{gintc}
\en
\endnumparts
where $|\phi_k(d)\rangle$ are given by \eref{deriv} and $d_u$ is defined in \eref{dxyz}. 
The intrinsic Cartesian components of \eref{gintc} vanish owing to the assumed
symmetries of $|\phi(d)\rangle$ discussed in \sref{TS}.
The weight appearing in the Bohr Hamiltonian of \eref{collham} is, according to \eref{detmat}, equal to
\beq\label{sqg}
\sqrt{g(d)}= \sqrt{(g_{00}(d)g_{22}(d)-g_{02}^2(d))g_{xx}(d)g_{yy}(d)g_{zz}(d)}.
\eeq
The intrinsic Cartesian components of matrix $\mt{f}$ inverse to $\mt{g}$ and its square root matrix $\mt{n}$ which appear in \eref{invinertcoll} and
\eref{potcoll} can be calculated from \eref{ginta} and \eref{gintb} according to \eref{sqmatintr} and \eref{invmatintr}.

In the Gaussian approximation the energy kernel $\mathcal{H}(\bls{\al},\bls{\al}')$ is given by \eref{gaussen}.
The zero order term in the expansion is
\beq\label{vdef}
v(\bls{\al})=v(d)=\frac{\mathcal{H}(\bls{\al},\bls{\al})}{\mathcal{I}(\bls{\al},\bls{\al})}=\langle\phi (d)|\hat{H}|\phi (d)\rangle .
\eeq
The second order term is defined by the matrix of the second order partial derivatives of quotient $\mathcal{H}/\mathcal{I}$ at
$\bls{\gm}=0$ as follows:
\beq\label{hdef}
   h^{\mu\nu}(\bls{\bta})=-\frac{\partial^2}
{\partial\gm_\mu\partial\gm_\nu}\left(\frac{\mathcal{H}(\bls{\bta},\bls{\bta})}
{\mathcal{I}(\bls{\bta},\bls{\bta})}\right),
\eeq
where the notation of \eref{firstder} is used. Due to \eref{norm0}, \eref{firstder} and \eref{gdef} we have:
\bn\label{hgdef}
h^{\mu\nu}(\bls{\bta})
& =&-\frac{\partial^2\mathcal{H}(\bls{\bta},\bls{\bta})}
{\partial\gm_\mu\partial\gm_\nu}+\frac{\partial^2\mathcal{I}(\bls{\bta},\bls{\bta})}
{\partial\gm_\mu\partial\gm_\nu}\mathcal{H}(\bls{\bta},\bls{\bta}) \nonumber \\
& =&-\frac{\partial^2\mathcal{H}(\bls{\bta},\bls{\bta})}
{\partial\gm_\mu\partial\gm_\nu}-g^{\mu\nu}(\bls{\bta})v(\bls{\bta}).
\en
After replacing the derivatives with respect to $\gm_{\mu}$ by the derivatives with respect to $\al_\mu$ $\mt{h}$ takes the following form
\bn\label{hdeffin}
  h^{\mu\nu}(\bls{\al})&=&\half\left[\langle\Phi^\mu (\bls{\al})|\hat{H}|\Phi^\nu (\bls{\al})\rangle 
-\langle\Phi (\bls{\al})|\hat{H}|\Phi^{\mu\nu} (\bls{\al})\rangle\right] \nonumber \\
&&-\langle\Phi^\mu (\bls{\al})|\Phi^\nu (\bls{\al})\rangle\langle\Phi (\bls{\al})|\hat{H}|\Phi (\bls{\al})\rangle .
\en
Differentiation with respect to $\al_\mu$ by means of \eref{dalpha} gives naturally $\mt{h}$ in the form of \eref{mat} with the following intrinsic Cartesian components which do not vanish:
\numparts
\bn\label{hint}
\fl h_{kl}(d)=\thalf [\langle\phi_k (d)|\hat{H}|\phi_l(d)\rangle -\langle\phi (d)|\hat{H}|\phi_{kl}(d)\rangle ]
-\langle\phi_k (d)|\phi_l(d)\rangle\langle\phi (d)|\hat{H}|\phi (d)\rangle \nonumber \\
 \mathrm{for}\; k,l=0,2, \label{hinta}\\
\fl h_{uu}(d)=\frac{1}{4d_u^2}\left[\langle\phi (d)|\hat{H}\hat{J}^2_u|\phi (d)\rangle
-\langle\phi (d)|\hat{J}^2_u|\phi (d)\rangle\langle\phi (d)|\hat{H}|\phi (d)\rangle\right] \nonumber \\
\quad \mathrm{for}\; u=x,y,z .\label{hintb}
\en
\endnumparts
The remaining intrinsic components vanish for the symmetry reasons (cf \sref{TS}).

We see that in order to calculate $v,\ \mt{g}$ and $\mt{h}$, and then the inverse inertial bitensor and the collective potential we should calculate
the mean values of the squares of angular momenta $\hat{J}^2_u$, Hamiltonian $\hat{H}$ and their products $\hat{J}^2_u\hat{H}$ within the deformation dependent intrinsic ground sate $|\phi (d)\rangle$, construct the states $|\phi_k(d)\rangle$ and $|\phi_{kl}(d)\rangle$ through differentiations of
$|\phi (d)\rangle$ with respect to the deformation parameters $d$ and calculate their overlaps and matrix elements of $\hat{H}$ within them.

\subsection{Inverse inertial functions and potential}\label{AV}

The intrinsic Cartesian components of the inverse inertial bitensor \eref{invinertcoll} do not depend on
the Euler angles $\omega$ and are  functions
of deformation $d$ only. We refer to the corresponding intrinsic components $A_{ab}(d),\ (a,b=0,2,x,y,z)$ of  matrix $A_{\mu\nu}(\bls{\al})$ as the inverse inertial functions.
It results from \eref{invinertcoll} using \eref{Dmu}, \eref{mat} and \eref{invmatintr} that the inverse inertial vibrational functions are equal to
\numparts
\bn\label{avib}
\fl A_{00}(d)=\frac{g_{22}^2(d)h_{00}(d)-2g_{02}(d)g_{22}(d)h_{02}(d)+g_{02}^2(d)h_{22}(d)}{(g_{00}(d)g_{22}(d)-g^2_{02}(d))^2}, \label{a00}\\
\fl A_{02}(d)=\frac{(g_{00}(d)g_{22}(d)+g_{02}^2(d))h_{02}(d)-g_{02}(d)g_{22}(d)h_{00}(d)-g_{02}g_{00}(d)h_{22}(d)}{(g_{00}(d)g_{22}(d)-g^2_{02}(d))^2},
\label{a02}\\
\fl A_{22}(d)=\frac{g_{02}^2(d)h_{00}(d)-2g_{02}(d)g_{00}(d)h_{02}(d)+g_{22}^2(d)h_{22}(d)}{(g_{00}(d)g_{22}(d)-g^2_{02}(d))^2} \label{a22}
\en
and the inverse inertial rotational functions read
\beq\label{auu}
A_{uu}(d)=\frac{h_{uu}(d)}{g_{uu}^2(d)}
\eeq
for $u=x,\ y,\ z$.
\endnumparts
Hence, the moments of inertia by definition (cf \cite{Pro09}) are
\bn\label{mominert}
\fl I_u(d)=\frac{4d_u^2}{A_{uu}(d)}=\frac{\langle\phi (d)|\hat{J}^2_u|\phi (d)\rangle^2}{\langle\phi (d)|\hat{H}\hat{J}^2_u|\phi (d)\rangle
-\langle\phi (d)|\hat{J}^2_u|\phi (d)\rangle\langle\phi (d)|\hat{H}|\phi (d)\rangle}.
\en
They resemble the Yoccoz moment of inertia \cite{Pei57,Yoc57} and are in form similar to those of a rigid body \cite{Une76}.

 Being a quadrupole scalar, the collective potential  depends obviously on deformation $d$ only.
\Eref{potcoll} for the potential contains matrices $\mt{n}$ and $\mt{a}$. Their intrinsic Cartesian components can be calculated from \eref{invmatintr},
\eref{sqmatintr} and \eref{a}.
It is worth noting that the zero-point energy corrections associated with the rotational modes come not only from the rotational 
components (of type $a_{uu},\ u=x,\ y,\ z$) of the matrices involved in the corresponding formula for the potential and the derivatives of these matrices with respect to $d$
but also from the derivatives with respect to $\omega$ coming from the differentiation of \eref{dalpha} with respect to $\al_\mu$. These latter derivatives
can be performed using \eref{mat} and \eref{lond}. Obviously, the final result for the potential does not depend on the Euler angles.

\section{Conclusions}\label{C}
Using the GCM we have generated the quadrupole collective excitations from the deformation-dependent intrinsic ground state which possesses
the $D^T_{2h}$ symmetry. To consider not only the quadrupole vibrations but also the quadrupole rotations the intrinsic state is rotated and
the three Euler angles of the rotation are attached to the two deformation parameters and form together the real electric quadrupole tensor which
plays a role of a set of generator coordinates. It turns out convenient to use just the quadrupole tensor, the real components of which are the
Cartesian coordinates in the five-dimensional space, instead of the Euler angles and the deformation parameters. In this way, periodic coordinates and a complicated topology of the space 
are avoided. The local GOA has been applied to the overlap and energy kernels. In this case the integral Hill-Wheeler
equation can be reduced to the differential equation having the form of the eigenvalue equation for the Bohr Hamiltonian. The reduction has been 
performed using the Fourier analysis of the energy kernel. The square root of determinant of the matrix defining the overlap in the GOA (the metric tensor)
is the weight in the Bohr Hamiltonian. The exact potential and the inverse inertial bitensor contains the exponential function of the five-dimensional
Laplacian operator and therefore, can be calculated only in an approximation. The simplest approximation consists in replacing the exponential operator
by the first term of its expansion --- the unit operator. This approximation corresponds to the potential and the inverse inertial bitensor obtained by the usual collective approach to the GCM. The next terms of the expansion can be, of course, taken into account in our approach, however, without settling the issue of convergence of the procedure.  

When deriving the Bohr Hamiltonian, a transformation of the quadrupole variables must be performed, which transforms
the metric tensor into the unit matrix. The resulting new set of variables is defined well when the square root matrix of the metric tensor
is an irrotational field (cf \cite{Oni75}). Obviously, this condition is equivalent to the condition which means that  the Riemannian space with 
the metric tensor in question being the Euclidean space. The assumption about the flatness of the space is apparently made in  other papers on
the multi-dimensional GOA even if this is not mentioned there (cf \cite{Goz85a,Goz85b}). The final form of the Bohr Hamiltonian does not seem to depend on this assumption. From the brief report by Kamlah \cite{Kam73} it seems that the assumption is not necessary, however, the proof of this fact has not been given there. Is it thus a technical condition coming from the fact that we are not able to calculate multi-dimensional Gaussian integrals? Or is it an essential condition? 
A similar problem appears in a sense in the case of the Podolsky-Pauli quantization prescription. The form of the quantum operator is simply assumed.
However, if one would want to derive this form from the Schr\"odinger operator in the Euclidean space, one should assume that the new variables are
curvilinear variables also in the Euclidean space and the kinetic energy is simply proportional to the Laplacian in curvilinear coordinates. It would then mean that the classical inertial matrix represents the metric tensor of an Euclidean space. We are left with this open problem.

\ack
The author is deeply indebted to Jacek Dobaczewski and Leszek Pr\'ochniak for their kindly assistance at every stage of his work. He would like to thank
both of them for good advices, critical remarks, constructive suggestions and valuable discussions.

\appendix
\section{Semi-Cartesian Wigner functions}\label{SCWF}
The quadrupole tensors transform themselves under rotations by means of the rotation matrices or the Wigner functions 
$\mathcal{D}^2_{\mu\nu}(\omega )$ which depend on a set of the Euler angles defining the rotation \cite{BM69}. It is convenient to introduce
the following linear combinations of the Wigner functions \cite{Roh82}:
\bn\label{WCart}
D_{\mu 0}(\omega ) &=&\mathcal{D}^2_{\mu 0}(\omega ), \nonumber \\
D_{\mu 2}(\omega ) &=& \frac{1}{\sqrt{2}}(\mathcal{D}^2_{\mu 2}(\omega ) +\mathcal{D}^2_{\mu -2}(\omega ) ), \nonumber \\  
D_{\mu x}(\omega ) &=& \frac{\rmi}{\sqrt{2}}(\mathcal{D}^2_{\mu 1}(\omega ) +\mathcal{D}^2_{\mu -1}(\omega ) ),  \\   
D_{\mu y}(\omega ) &=& \frac{1}{\sqrt{2}}(\mathcal{D}^2_{\mu 1}(\omega ) -\mathcal{D}^2_{\mu -1}(\omega ) ), \nonumber \\  
D_{\mu z}(\omega ) &=& \frac{\rmi}{\sqrt{2}}(\mathcal{D}^2_{\mu 2}(\omega ) -\mathcal{D}^2_{\mu -2}(\omega ) ). \nonumber
\en
Let us call them the ''semi-Cartesian" Wigner functions. The complex conjugate functions are
\beq\label{Dcompl}
D^\mu_k(\omega ) = D_{\mu k}^\ast (\omega ) =(-1)^\mu D_{-\mu k}(\omega ).
\eeq
 The orthogonality conditions for the semi-Cartesian Wigner functions take the form:
\bn
\sum_{\mu=-2}^2D^\mu_k(\omega )D_{\mu l}(\omega ) &=& \delta_{kl}, \label{Dmu} \\
\sum_{k}D^\mu_k(\omega )D_{\nu k}(\omega ) &=& \delta^\mu_{\nu}. \label{Dk}
\en

The drift angular momentum operators $L_u(\om )$, $u=x,\ y,\ z$, act on the semi-Cartesian Wigner functions as follows \cite{Roh82}
\bn\label{lond}
&& L_x(\omega )D_{\mu 0}(\omega )= -\rmi\sqrt{3}D_{\mu x}(\omega ),\quad L_x(\omega )D_{\mu y}(\omega )= -\rmi D_{\mu z}(\omega ),
\nonumber \\
&& L_x(\omega )D_{\mu x}(\omega )= \rmi\sqrt{3}D_{\mu 0}(\omega )+\rmi D_{\mu 2}(\omega ), \nonumber \\
&& L_x(\omega )D_{\mu z}(\omega )= \rmi D_{\mu y}(\omega ), \quad L_x(\omega )D_{\mu 2}(\omega )= -\rmi D_{\mu x}(\omega ), \nonumber \\
&& L_y(\omega )D_{\mu 0}(\omega )= \rmi\sqrt{3}D_{\mu y}(\omega ),\quad   L_y(\omega )D_{\mu x}(\omega )= \rmi D_{\mu z}(\omega ),
\nonumber \\
&& L_y(\omega )D_{\mu y}(\omega )=- \rmi\sqrt{3}D_{\mu 0}(\omega )+\rmi D_{\mu 2}(\omega ),  \\
&& L_y(\omega )D_{\mu z}(\omega )=- \rmi D_{\mu x}(\omega ), \quad L_y(\omega )D_{\mu 2}(\omega )=- \rmi D_{\mu y}(\omega ),\nonumber \\
&& L_z(\omega )D_{\mu 0}(\omega ) =0, \quad  \nonumber \\
&& L_z(\omega )D_{\mu x}(\omega ) =\rmi D_{\mu y}(\omega ), \quad L_z(\omega )D_{\mu y}(\omega ) =-\rmi D_{\mu x}(\omega ), \nonumber \\
&& L_z(\omega )D_{\mu z}(\omega ) =2\rmi D_{\mu 2}(\omega ), \quad L_z(\omega )D_{\mu 2}(\omega ) =-2\rmi D_{\mu z}(\omega ). \nonumber
\en
It is seen that all the three components of the drift angular momentum change the Cartesian indeces of the Wigner functions.

\section{Symmetric matrices as isotropic functions of the quadrupole tensor}\label{QM}
Any symmetric quadrupole bitensor (symmetric matrix $5\times 5$) $m_{\mu\nu}(\al )$ which is an isotropic real function of 
the quadrupole tensor $\bls{\al}$
can be determined either by six scalar functions of $d=(d_0,\ d_2)$ or by the six intrinsic Cartesian components in the following way (cf \cite{Pro09}):
\beq\label{mat}
m_{\mu\nu}(\bls{\al} )=\sum_{k,l}D_{\mu k}(\omega )D_{\nu l}(\omega )m_{kl}(d),
\eeq 
where the Cartesian indeces $k,\ l$ run over symbols $0,\ 2,\ x,\ y,\ z$. The Euler angles $\omega$ determine the orientation of the intrinsic system.
The Cartesian matrix $m_{kl}(d)$ is real symmetric and has the following structure:
\bn\label{ intrmat}
\mt{m}(d) &=&\left(\ba{ccccc}m_{00}(d) & m_{02}(d) & 0 & 0 & 0 \\
                                     m_{02}(d) & m_{22}(d) & 0 & 0 & 0 \\
                                      0 & 0 & m_{xx}(d) & 0 & 0 \\
                                      0 & 0 & 0 & m_{yy}(d) & 0 \\
                                      0 & 0 & 0 & 0 &m_{zz}(d) \ea\right) .
\en
The determinant of $\mt{m}$ reads
\beq\label{detmat}
m=\det{(\mt{m})}= (m_{00}m_{22}-m_{02}^2)m_{xx}m_{yy}m_{zz}=m_v^2 m_{xx}m_{yy}m_{zz}.
\eeq
The Cartesian components of matrix $\mt{m}(\bls{\al})$ are:
\beq\label{cart}
m_{kl}(\bls{\al})= C^\mu_kC^\nu_lm_{\mu\nu}(\bls{\al} ),
\eeq
where $C^\mu_k=D^\mu_k(\om =0)$.

If matrix $m_{\mu\nu}$ is not singular i.e. $m\ne 0$ matrix $(\mt{m}^{-1})_{\mu\nu}$ inverse to $m_{\mu\nu}$ can be defined as
\beq\label{invmat}
m_{\mu\kappa}(\bls{\al} )(\mt{m}^{-1})^{\kappa\nu}(\bls{\al} )=\delta^\nu_\mu .
\eeq
The intrinsic Cartesian matrix inverse to $\mt{m}$ is:
\bn\label{invmatintr}
\fl  \mt{m}^{-1}(d) \nonumber \\
\fl = \left(\ba{ccccc}m_{22}(d)/m^{2}_v(d) & -m_{02}(d)/m^{2}_v(d)& 0 & 0 & 0 \\
                                                   -m_{02}(d)/m^{2}_v(d) & m_{00}(d)/m^{2}_v(d) & 0 & 0 & 0  \\
                                                    0 & 0 & 1/m_{xx}(d) & 0 & 0 \\
                                                     0 & 0 & 0 & 1/m_{yy}(d) & 0 \\
                                                       0 & 0 & 0 & 0 & 1/m_{zz}(d) \ea\right) .
\en

 If  $\mathsf{m}$ is positive definite i.e.  $m_{00}$, $m_{22}$, $m_{xx}$, $m_{yy}$, $m_{zz}$ and $m_v^2=m_{00}m_{22}-m_{02}^2$ 
are all positive
 the positive-definite square root matrix $(\mt{m}^{1/2})_{\mu\nu}$ such that
\beq\label{sqmat}
m^{\mu}_{\nu}(\bls{\al} ) = (\mt{m}^{1/2})^{\mu}_{\kappa}(\bls{\al} )(\mt{m}^{1/2})^{\kappa}_{\nu}(\bls{\al} )
\eeq
can be defined. It is real symmetric and has the structure of \eref{mat} like $\mt{m}$. The nonvanishing entries of 
the intrinsic Cartesian matrix $(\mt{m}^{1/2})_{kl}$ are related
to components $m_{kl}$ as follows:
\bn\label{sqmatintr}
(\mt{m}^{1/2})_{00} &=&\frac{m_{00}+m_v}{\sqrt{m_{00}+m_{22} +2m_v}}, \nonumber \\
(\mt{m}^{1/2})_{22} &=&\frac{m_{22}+m_v}{\sqrt{m_{00}+m_{22} +2m_v}}, \nonumber \\
(\mt{m}^{1/2})_{02} &=&\frac{m_{02}}{\sqrt{m_{00}+m_{22} +2m_v}}, \nonumber \\
(\mt{m}^{1/2})_{xx} &=& \sqrt{m_{xx}}, \quad (\mt{m}^{1/2})_{yy} = \sqrt{m_{yy}}, \quad (\mt{m}^{1/2})_{zz} = \sqrt{m_{zz}}.
\en

\end{document}